\documentclass[aps,prl,twocolumn,groupedaddress]{revtex4}
\usepackage[draft]{hyperref}
\usepackage{amsmath,amssymb}
\usepackage{graphicx}
\usepackage{color}

\newcommand{\beque}{\begin{equation*}}
\newcommand{\eeq}{\end{equation}}
\newcommand{\beq}{\begin{equation}}
\newcommand{\eeque}{\end{equation*}}
\newcommand{\beqnl}{\begin{eqnarray}}
\newcommand{\eeqna}{\end{eqnarray*}}
\newcommand{\beqna}{\begin{eqnarray*}}
\newcommand{\eeqnl}{\end{eqnarray}}

\begin{document}

\title{Blackbody emission from light interacting with an effective moving dispersive medium.}

\author{M.~Petev,$^1$ N.~Westerberg,$^1$ D.~Moss,$^1$ E.~Rubino,$^{2}$ C.~Rimoldi,$^{2}$  S.L.~Cacciatori,$^{2}$  F.~Belgiorno,$^{3}$ and  D.~Faccio$^{1,*}$}

\address{
$^{1}$School of Engineering and Physical Sciences, SUPA, Heriot-Watt University, Edinburgh EH14 4AS, UK\\
$^{2}$Dipartimento di Scienza e Alta Tecnologia, Universit\`a dell'Insubria, Via Valleggio 11, IT-22100 Como, Italy \\
$^{3}$Dipartimento di Matematica, Politecnico di Milano, Piazza Leonardo 32, 20133 Milano, Italy
}
\email{d.faccio@hw.ac.uk}

\begin{abstract}
Intense laser pulses excite a nonlinear polarisation response that may create an effective flowing medium and, under appropriate conditions, a blocking horizon for light. Here we analyse in detail the interaction of light with such laser-induced flowing media, fully accounting for the medium dispersion properties. An analytical model based on a first Born-approximation is found to be in excellent agreement with numerical simulations based on Maxwell's equations and shows that when a blocking horizon is formed, the stimulated medium scatters light with a blackbody emission spectrum. Based on these results, diamond is proposed as a promising candidate medium for future studies of Hawking emission from artificial, dispersive horizons. 
\end{abstract}

%\pacs{47.55.dp, 43.25.Yw, 47.55.dd, 52.50.Jm}

\maketitle

Recent developments in the understanding of light propagation have led to evidence that by using intense laser pulses propagating in a nonlinear medium, it is possible to create an effective medium that flows with the same speed as the laser pulse, i.e. at speeds close to or, as a consequence of dispersion, even higher than the speed of light at other frequencies in the medium \cite{ulf,faccioCP,faccioEPL,carusottoPRA,belgiornoPRD,cacciatoriNJP,visser_review,mendonca,superluminal}. Indeed, in a medium with a third order (also called ``Kerr'') nonlinear polarisation response the refractive index of the medium is given by $n=n_0+n_2I(z-vt)$, where $n_0$ is the background index, $n_2$ is the  nonlinear Kerr index and $I(z-vt)$ is the laser pulse intensity profile, travelling along the $z$ direction with velocity $v$. In a typical condensed medium, e.g. glass, the maximum amplitude of the laser pulse induced refractive index perturbation is $\delta n_\textrm{max}= n_2I_\textrm{max}\sim0.01-0.001$. This is sufficient to scatter light, e.g. either light from the laser pulse itself (self-scattering process) or from a second weak probe pulse (induced scattering process).\\
Under appropriate conditions one may also potentially observe induced scattering (and excitation) of photons that originate from the vacuum state. One motivation for studying such an effect  lies in the prediction that the $\delta n$ may be described in terms of a horizon that mimics the event horizon of gravitational black hole \cite{ulf,faccioCP,faccioEPL,carusottoPRA,belgiornoPRD,cacciatoriNJP,visser_review}. Indeed, in the absence of material dispersion, it is possible to write the effective spacetime metric for the moving $\delta n$ as $ds^2=c^2dt^2+(dr-Vdt)^2$. This is equivalent to the P{ainlev\'e-Gullstrand metric showing that in the reference frame of the $\delta n$ the medium is effectively flowing with speed $V$, that is a Galilean velocity ($-\infty<V<+\infty$) and is determined by $n$ and $v$. A horizon is formed when $V=c$ \cite{river,belgiornoPRD}. Interaction with the vacuum state leads to the emission of photon pairs that are analogous to Hawking emission from a black hole. Recent measurements claimed the observation of a spontaneous emission from a laser pulse induced moving $\delta n$ that appeared to have  some of the features predicted for the analogue Hawking emission \cite{belgiornoPRL,rubinoNJP}. However, these measurements are not considered as conclusive and further evidence is required \cite{PRLcomment,visserdn,unruhdn}. \\
In this Letter we investigate in detail the interaction between a  $\delta n$ moving at a generic speed, $v$, and a probe light pulse. Our analytical and numerical models fully account, without any approximations for material dispersion, the full shape of the $\delta n$ and allow a comprehensive analysis of the interaction dynamics for varying $\delta n$ speed. 
The input probe pulse is  scattered into two output modes one with positive and the other with negative comoving frequencies \cite{NRR}. A remarkable feature of this stimulated emission is that the negative mode emissivity is dictated by a blackbody law with a temperature that is directly related to the steepness of the moving $\delta n$. This prediction is also fully confirmed  analytically within a Born approximation model that provides physical insight into the scattering mechanism. 
The blackbody emission is predicted to occur for $\delta n$ speeds such that the light pulse is slowed to the point that it cannot traverse the $\delta n$ - the $\delta n$ has a  blocking horizon.  It is in this regime that we propose a setting for future experiments in diamond that exhibits the required conditions to observe analogue Hawking emission.\\
\begin{figure}[t]
\centering
\includegraphics[width=8cm]{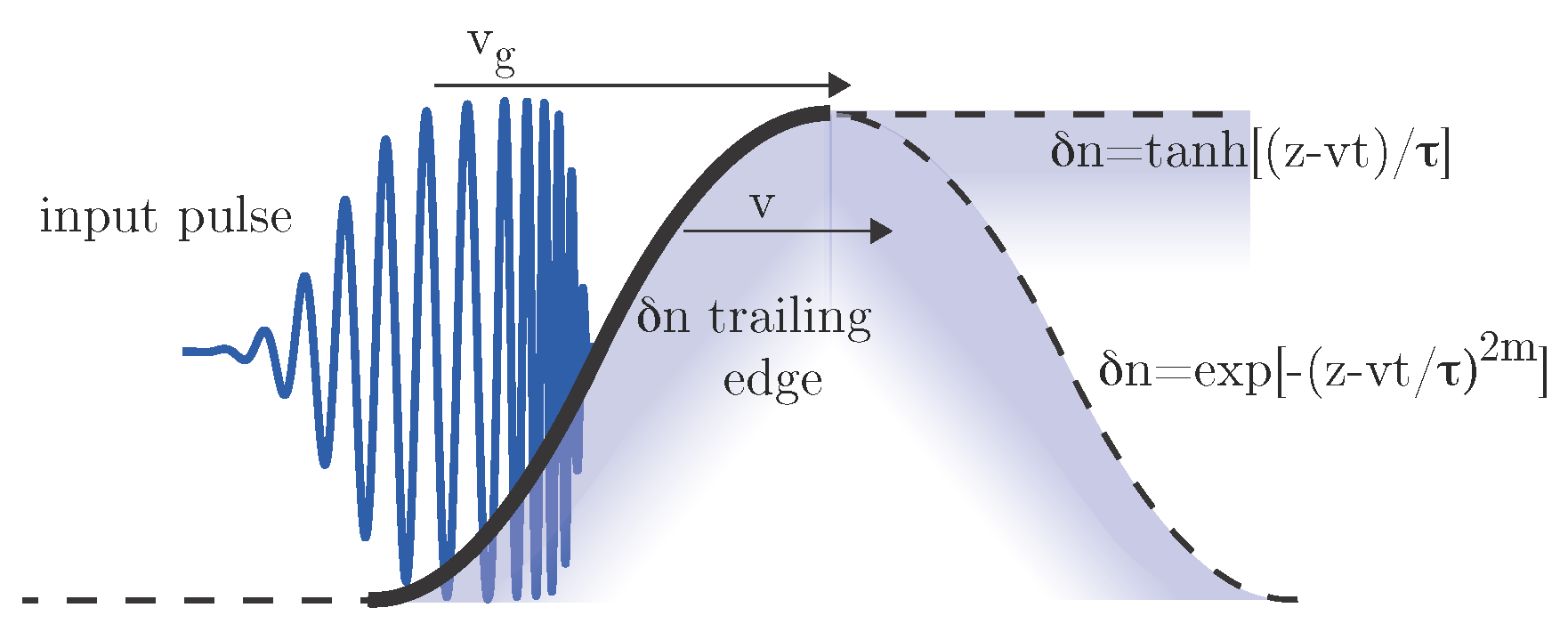}
\caption{Schematic drawing of interaction geometry: the input light pulse propagates with speed $v_g$, catches up and interacts with the trailing edge (think solid curve)  of the moving $\delta n$, propagating with speed $v\lesssim v_g$. }
\label{fig0}
\end{figure}
{\emph{Scattering from a moving $\delta n$:}} A soliton propagating in a dispersive medium will shed light through a mechanism known as resonant (or dispersive wave) radiation (RR) \cite{wai,cerenkov,skryabin,dudley}. This emission may be described as a self-induced scattering process whereby light from the soliton is scattered into a frequency-shifted mode by the self-induced Kerr $\delta n$. It was recently shown that a second scattered mode also exists: this mode has negative frequency in the reference frame comoving with the soliton and has been called ``negative-frequency resonant radiation'' (NRR) \cite{NRR}. Both these modes are found within the framework of a  model that neglects all nonlinear effects and simply considers the soliton $\delta n$ and models how light is scattered within the first Born-approximation \cite{SR}. The model therefore generalises RR and NRR generation beyond soliton physics to include also systems in that do not support solitons (e.g. a 3-dimensional pulse with transverse spatial dynamics or an intense pulse in the normal group velocity dispersion regime). The scattered amplitudes of the two modes are given by \cite{SR}
\begin{eqnarray}
S(\omega_{\rm{RR}})  = i \frac {v\omega_{\rm{RR}}^2 n_0}{2c^2 k_z(\omega_{\rm{RR}})} e^{i\frac{(\omega_{\textrm{RR}}-\omega_{\textrm{IN}})z}{v}} \hat R (\omega_{\textrm{RR}}-\omega_{\textrm{IN}}) \label{eq:Nborn1}\\
S(\omega_{\rm{NRR}}) =  i \frac {v\omega_{\rm{NRR}}^2 n_0}{2c^2 k_z(\omega_{\rm{NRR}})} e^{i\frac{(\omega_{\rm{NRR}}+\omega_{\rm{IN}})z}{v}}\hat{R}(\omega_{\rm{NRR}}+\omega_{\rm{IN}})
\label{eq:Nborn2}
\end{eqnarray}
where $\hat{R}$ is the $\delta n$ Fourier transform and $\omega_{\rm{RR,NRR,IN}}$ are the RR, NRR and input probe pulse frequencies.\\
In the following we present a series of numerical simulations and a comparison with predictions based on the Born approximation equations \eqref{eq:Nborn1}  and \eqref{eq:Nborn2} with which we estimate the amplitudes of the RR and NRR waves generated by an input probe pulse interacting with a moving $\delta n$, for varying frequency of the input pulse. The system considered is depicted in Fig.~\ref{fig0} - a probe light pulse interacts with the trailing edge of a moving $\delta n$. The $\delta n$ may thus be modelled using any function that describes a rising refractive index front (two examples are shown in the figure). Backward propagating modes are neglected on the basis that their overlap times with the input pulse and $\delta n$ are extremely short and they will therefore carry negligible energy. The numerical simulations were carried out using both FDTD \cite{FDTD} and the UPPE \cite{UPPE,SR}  algorithms all under the same input conditions. Both methods directly solve Maxwell's equations however the FDTD algorithm supports backward propagating modes whilst the UPPE does not. There were no discernible differences between the outputs of these two codes hence lending further support to neglecting the backward modes. In Fig.~\ref{fig1} we show examples obtained with the FDTD code with $\delta n = \delta n_\textrm{max}\exp{[-[(z-vt)^2/\sigma^2]^m]}$ where $\delta n_\textrm{max}=0.01$, $\sigma$ and $m$ are chosen so that the $\delta n$ front has a 7 fs rise time and also such that this rise time is much shorter than the overall width of the $\delta n$, thus ensuring that light only interacts with the trailing edge. The dispersion is chosen so as to resemble that of diamond and is shown in the figure in $(\omega^\prime,\omega)$ frequency coordinates, where primed quantities indicate that they refer to the comoving reference frame, $\omega^\prime=(\omega-vk)$, where $k$ is the light pulse wave-vector.\\
\begin{figure}[t]
\centering
\includegraphics[width=8.5cm]{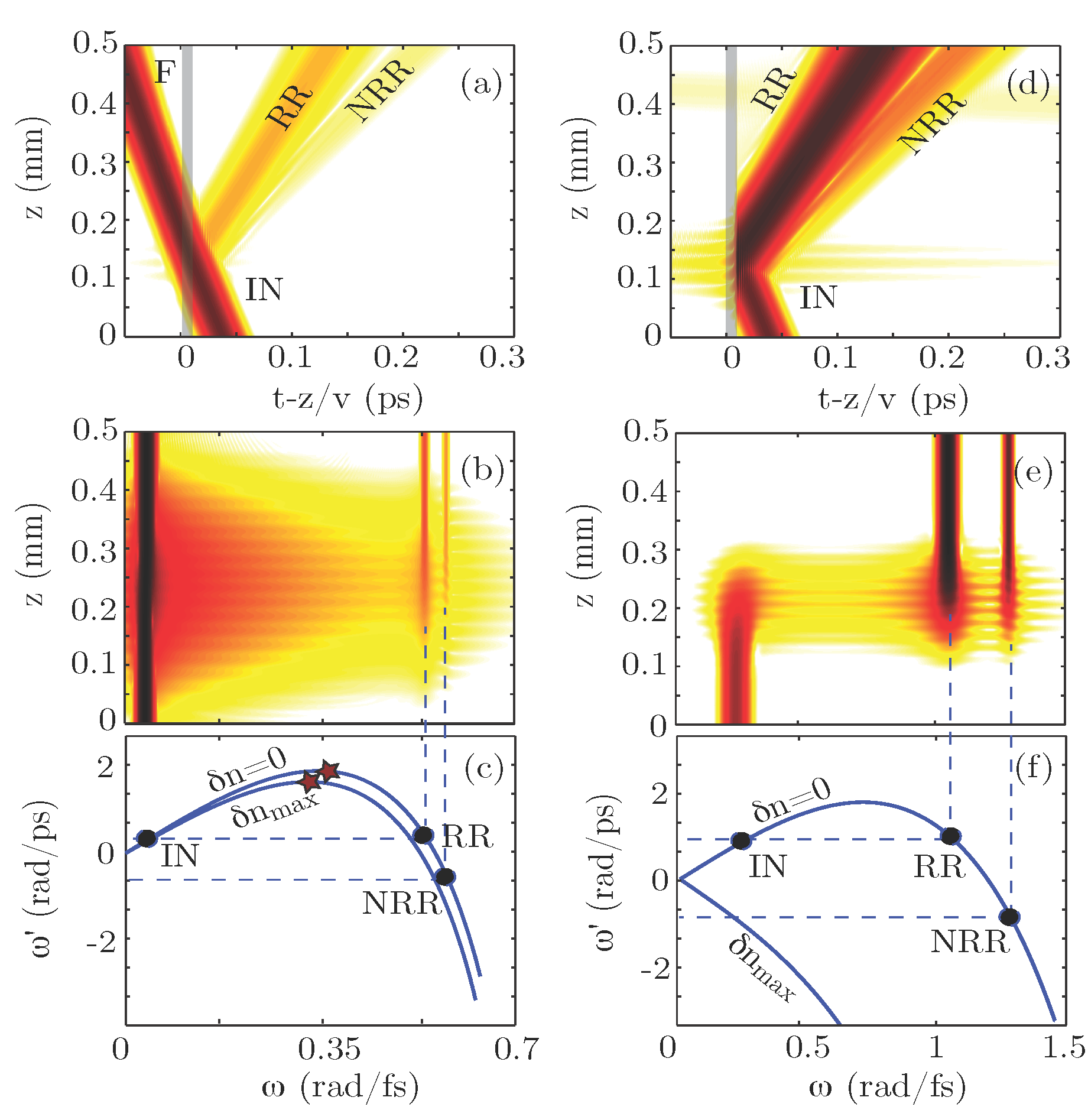}
\caption{Example of a numerically simulated interaction of a probe pulse with a moving $\delta n$. (a), (b) and (c) show the temporal envelope profile evolution, spectral evolution and relevant dispersion curves, respectively, for the case in which the $\delta n$ moves too slowly to form a blocking horizon. (d), (e) and (f) show similar figures for a faster $\delta n$ such that a blocking horizon is formed. The $\delta n$ rising front is shown as a grey shaded area in (a) and (d). All intensity plots are shown over 4 decades in logarithmic scale.}
\label{fig1}
\end{figure}
Figure~\ref{fig1}(a) shows the  temporal envelope profile of an input $5$ $\mu$m wavelength probe beam interacting with the $\delta n$ in the comoving frame for the generic case in which the $\delta n$ speed, $v=1.9\times10^8$ m/s is too slow to actually completely block the impinging light for $\delta n_\textrm{max}=0.01$. We refer to this as the ``non-blocking'' case: the input light will be transmitted through the $\delta n$ into a mode indicated with $F$ and will only be partly scattered into the RR and NRR modes that are reflected  backwards in the comoving frame (but are  travelling forwards in the laboratory frame). This partial conversion is also clear in the frequency spectrum evolution in Fig.~\ref{fig1}(b)  that shows generated peaks that are in excellent agreement with the  frequencies predicted from momentum conservation, i.e. comoving frequency $\omega^\prime$ conservation shown as the horizontal dashed lines intersecting the dispersion relation in Fig.~\ref{fig1}(c). Generalised Manley-Rowe (photon number balance) relations may be derived for an effective moving medium that, accounting for the modes present in this case read as $|F|^2+|RR|^2-|NRR|^2=1$ where $|\cdot|$ indicates the mode photon number normalized with respect to the input mode  (see SM).\\
Figures~\ref{fig1}(c), (d) and (e) show similar results but now $v=2.07\times10^8$ m/s has been increased such that the input light pulse is slowed down upon interacting with the $\delta n$ rising front to the point that its group velocity becomes equal to $v$, i.e. the $\delta n$ presents a ``blocking horizon'' for light. In this case the input mode is completely converted to the RR and NRR modes and $|RR|^2-|NRR|^2=1$.\\
We now examine the ratio $r=|NRR|^2/|RR|^2$ for varying input frequency. In Fig.~\ref{fig2}(a) we show a typical example for the blocking case in logarithmic scale (red points are results from numerical simulations the solid line is a linear fit). Similar results are found for all values of $v$, i.e. also in the non-blocking case. The surprising feature here is that $r$ has a clear exponential dependence over more than 8 decades. If we then combine this exponential dependence $r=\exp(-\alpha\omega')$ with the generalised Manley-Rowe relations we find that $|NRR|^2=(1-|F|^2)/[\exp(\alpha\omega')-1]$. In the blocking case ($|F|^2=0$), this implies that the NRR mode emission follows a blackbody law with temperature given by $T=\hbar/(k_\textrm{B}\alpha)$. The NRR photon number will thus diverge close to $\omega^\prime=0$ as shown in Fig.~\ref{fig2}(a) (solid line indicated with ``NRR''). The moving $\delta n$ is therefore effectively transforming the input light, regardless of its state, into an output NRR mode that has a comoving blackbody spectrum. 
%%%%%
%
\begin{figure}[t]
\centering
\includegraphics[width=8.5cm]{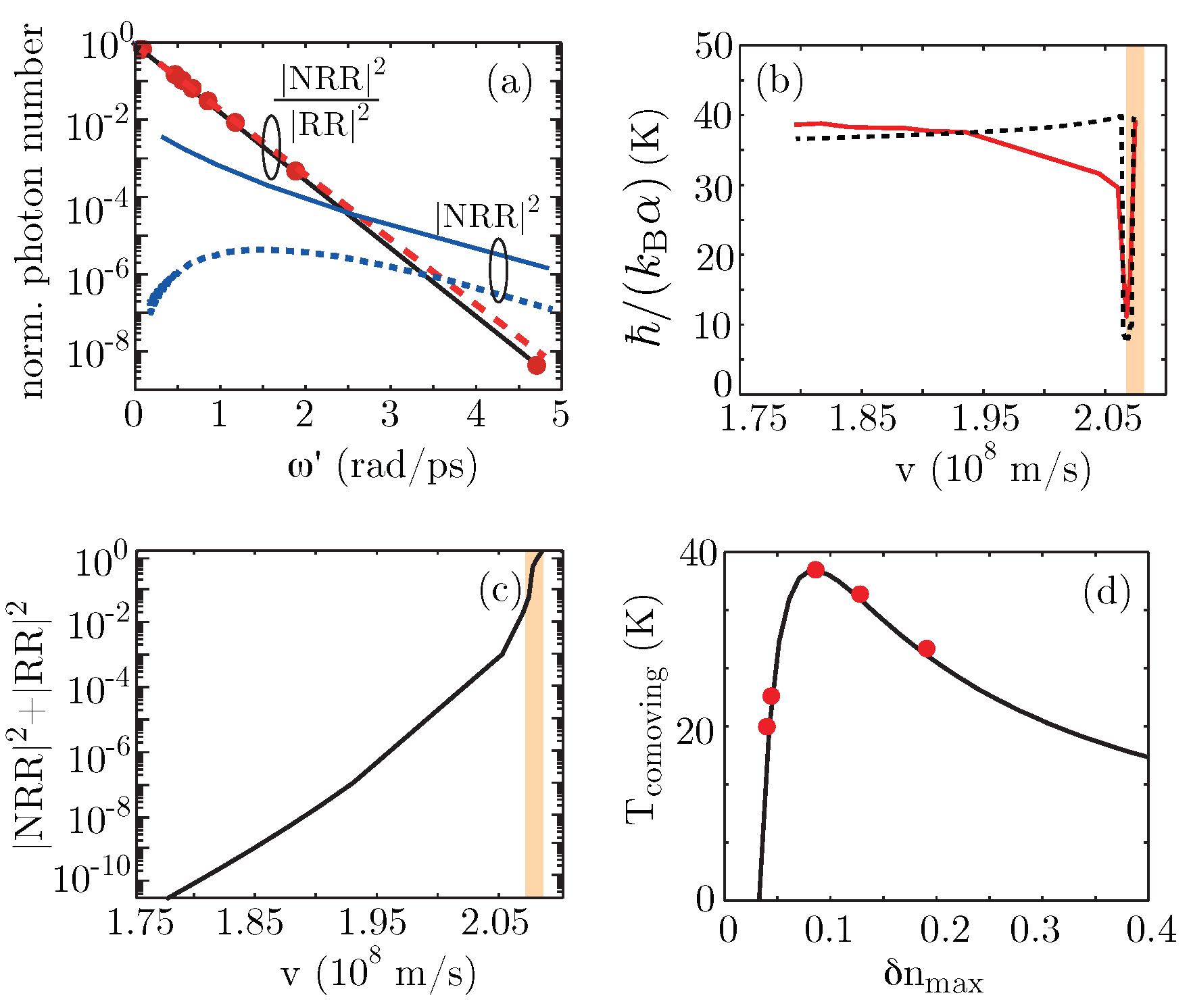}
\caption{(a) Numerical simulation for $v=1.9\times10^8$ m/s of $r=|NRR|^2/|RR|^2$ for varying input frequency (dots) and best fit with exponential function (solid line). Dashed line - Born approximation calculation. Also shown is the normalized photon number $|NRR|^2$ for the blocking (solid blue line) and non-blocking (dotted line) case. (b) $\hbar/k_\textrm{B}\alpha$ [derived from graphs as in (a)] for varying $\delta n$ speed -  simulations (solid line) and Born approximation model (dashed line). The shaded area indicates speeds for which a blocking horizon is formed. (c) Total normalised photon count for varying speed. (d) Comparison between numerically estimated emission temperature, $T=\hbar/k_\textrm{B}\alpha$ (dots) and theoretical Hawking emission temperature estimated from the $\delta n$ gradient (solid line) for varying maximum index change, $\delta n_\textrm{max}$ (blocking case).}
\label{fig2}
\end{figure}
We note that we can also evaluate $r$ directly from the first Born approximation relations, Eqs.~\eqref{eq:Nborn1}-\eqref{eq:Nborn2} and the same exponential dependence for $r$ is found [shown as a dashed line in Fig.~\ref{fig2}(a)]. For a wide range of step-like functions, the Born approximation can even be solved exactly and thus proves the robustness of this result (see SM).\\ 
 In Fig.~\ref{fig2}(b) we show the $\hbar/k_\textrm{B}\alpha$ obtained from similar curves as shown in Fig.~\ref{fig2}(a) for varying $\delta n$ speeds: the solid line is the result from numerical simulations, the dashed line shows the Born approximation result and the two curves are in overall good  agreement.
 However, we note that in crossing over into the non-blocking case the appearance of the additional transmitted mode ($|F|^2>0$) significantly distorts the blackbody spectrum. Indeed, if we plot the NRR photon number for a non-blocking case [dotted line in Fig.~\ref{fig2}(a)] we see that at small frequencies the mode number drops to zero rather than diverging as would be expected for a 1D blackbody emission. This is in keeping with quantum calculations in a generic `flowing fluid' setting \cite{robertson} and can be understood by noting that for high enough frequencies a blocking horizon is still present [region between the dispersion curve maxima indicated with stars in Fig.~\ref{fig2}(c)] but low frequencies are not blocked at all, implying that the interaction time with the $\delta n$ trailing edge is significantly reduced and any scattering process is suppressed. This suppression is also observed in Fig.~\ref{fig2}(c) that shows the total photon number  $|NRR|^2+|RR|^2$ evaluated from the numerical simulations. As can be seen, this number varies by several orders of magnitude with varying $v$ and is dramatically enhanced only in the presence of a blocking horizon. \\
\begin{figure}[t]
\centering
\includegraphics[width=8cm]{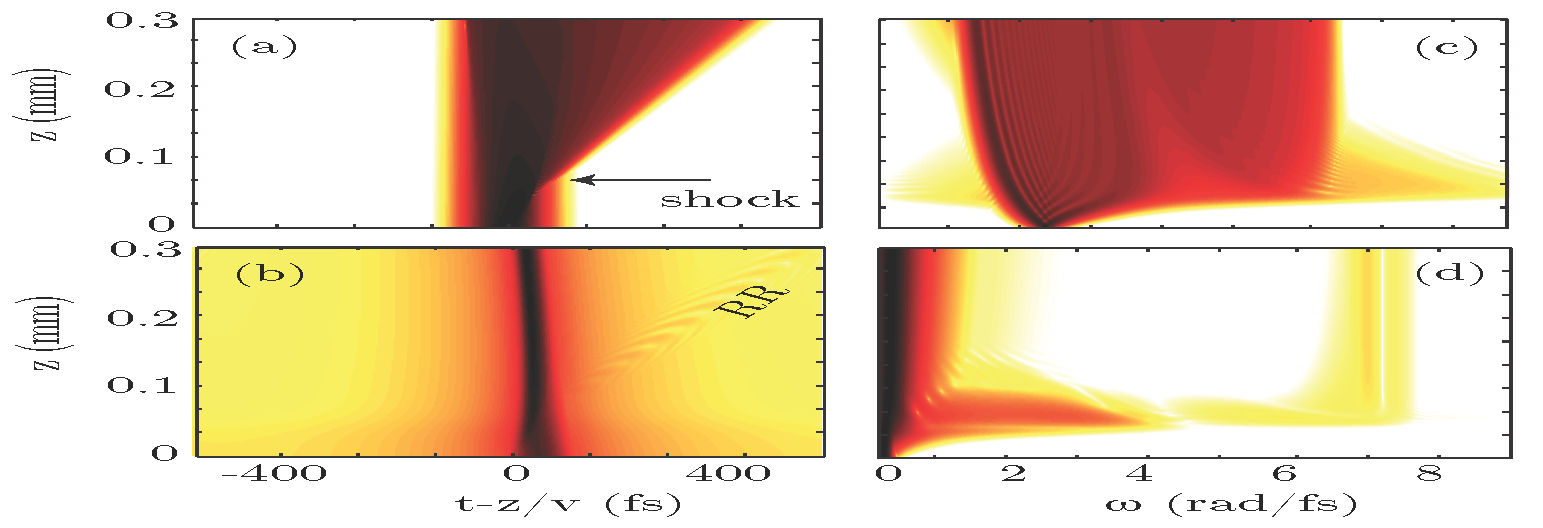}
\caption{ Numerical simulation of the nonlinear propagation of an 800 nm pump pulse and a co-propagating 15 THz probe pulse  (shown in logarithmic scale over 4 decades). Temporal profile evolution of the 800 nm pump pulse (a), and of the THz pulse, (b). The arrow indicates the propagation distance at which a shock front forms on the pump pulse. ``RR'' indicates emission from the THz pulse scattered from the pump shock front. (c)  shows the pump pulse spectral evolution. The THz pulse spectrum (6 decades in logarithmic scale) in (d) shows a clear peak centred around the $\omega'=0$ point, corresponding to a 7 rad/fs lab frame frequency.  }
\label{fig3}
\end{figure}
We note that a blocking horizon may be likened to a white hole horizon which in turn has been predicted to lead to Hawking emission \cite{unruh,carusottoPRA,ulf,belgiornoPRD,visser_review}, i.e. emission of a blackbody spectrum.
In the presence of a blocking horizon we may therefore evaluate the Hawking temperature according to $T=(1/2\pi)\gamma^2 v |dn/dt|$ (evaluated at the horizon) \cite{belgiornoPRD}. In Fig.~\ref{fig2}(d) we compare the (comoving) Hawking temperature (solid curve) and the numerically estimated temperatures (red dots) for $\delta n$ with a fixed $v$ and varying $\delta n_\textrm{max}$. As can be seen, the agreement is excellent over a wide range of values, including for $\delta n_\textrm{max}$ values that are experimentally accessible through the nonlinear Kerr effect.\\
We note that in the blocking case, the Manley-Rowe relations imply that $|RR|^2+|NRR|^2>1$, i.e. the photon numbers are amplified (at the expense of the moving $\delta n$) \cite{SR}. Therefore, if the input probe pulse were to be reduced to the level of the quantum noise fluctuations, then the results shown above predict the spontaneous emission of a blackbody spectrum. In other words, the noise of the amplifier is characterised by a temperature that remains constant across all frequencies. This is therefore a somewhat different kind of amplifier with respect to better known examples in optics, e.g. optical parametric amplifiers that have a noise temperature that scales linearly with frequency, $T=\hbar\omega/k_\textrm{B}$ \cite{seigman}. Moreover, the link with gravitational horizons, implies that the amplifier noise measured in an actual experiment may be likened to spontaneous Hawking radiation. The blackbody dependence implies that the amplifier gain scales as $1/\omega'$ for $\omega'\rightarrow 0$. In the comoving frame light is coupled between modes that have a constant or nearly constant $\omega^\prime\sim 0$. Looking at Fig.~\ref{fig1}(f) we see that if the input mode has lab. frequency $\omega$ close to zero then the output mode will appear at high frequencies, typically in the UV region,  in the laboratory frame. The amplifier therefore converts radiation from the low to the high (lab frame) frequency modes and the $1/\omega'$ blackbody divergence, i.e. Hawking emission  will appear as a divergence (or peak) centred at the UV lab frame frequency. \\
The high amplification gain when coupling between $\omega^\prime\sim 0$ modes also indicates possible methods for efficiently observing these scattering effects and possible applications. 
For example, in Fig.~\ref{fig3} we show numerical simulations of a weak 15 THz probe (20 $\mu$m wavelength) pulse interacting with  a Ti:Sapph laser pulse with 800 nm wavelength, 60 fs duration and input intensity 80 TW/cm$^2$, propagating in a 500 $\mu$m think diamond sample. These simulations were performed using two  UPPE equations, one for the pump and one for the THz pulse that are coupled only through a nonlinear cross-phase modulation term in the THz pulse equation $\propto2n_2I$ (details of the code can be found e.g. in Ref.~\cite{SR}). The self-phase modulation term $\propto n_2I$ is included in the 800 nm pump equation but four wave mixing and third harmonic generation have been purposely neglected. %We first consider a probe pulse with a 1 $\mu$m wavelength, slightly delayed with respect to the 800 nm pump pulse. It travels slightly faster than the pump pulse, catches up with it and then encounters a blocking horizon where it is completely frequency converted into the UV region. 
There is no blocking horizon for the probe pulse but nevertheless the spectrum clearly shows relatively efficient scattering to the UV that occurs predominantly when a shock front, i.e. the steepest gradient forms on the pump pulse [indicated by an arrow if Fig.~\ref{fig3}(a)] in agreement with Eqs.~\eqref{eq:Nborn1} and \eqref{eq:Nborn2}. Most interestingly, the output mode is peaked around the $\omega^\prime\sim0$, corresponding to a lab frequency of $7$ rad/fs (270 nm wavelength). In other words, laboratory reference frame frequencies in the THz or multi-THz region are already sufficiently low to excite the $1/\omega^\prime$ gain of the amplifier. This in turn provides indications of the spectral range of the noise fluctuations that may be spontaneously excited by the $\delta n$ and lead to a spontaneous (i.e. seeded by thermal or vacuum noise) emission peak in the UV. At room temperature, the background blackbody photon mode density at 15 THz is $\sim10^5$ and thus significantly larger that the quantum vacuum noise mode density equal to 1/2 a photon per mode: spontaneous emission will be seeded by the thermal background. However, it is sufficient to cool the diamond sample to $\sim 30$ K to invert the situation such that thermal fluctuations are dominated by more than 1 order of magnitude by quantum vacuum noise. \\
{\emph{Conclusions:}} a numerical and analytical evaluation based on the Born scattering approximation predicts that light is dramatically transformed by a moving medium. The  Born approximation model used here could also be extended to provide exact analytical relations and study scattering phenomena  in other systems in which linear waves interact with a moving, dispersive medium, e.g. gravity waves in water or acoustic oscillations in Bose-Einstein condensates. 
Optical horizons have been proposed for the measurement of Hawking emission but also for all-optical transitors \cite{demircan} but other applications of an effective moving medium  to be considered in future work could be relatively efficient THz detection by frequency conversion in diamond (even without a horizon, as shown in Fig.~\ref{fig3}) or generation of squeezed vacuum states in the UV region. \\
D.F. acknowledges discussions with I. Carusotto, S. Finazzi and financial support from the Engineering and Physical Sciences Research Council EPSRC, Grant EP/J00443X/1 and from the European Research Council under the European Union's Seventh Framework Programme (FP/2007-2013) / ERC Grant Agreement n. 306559.

%%%%%%%%%%%%%%%%%%%%%%%%%%%%%%%%%%

\section{Supplemental information}
In the following supplemental information we show results concerning the case of a superluminally moving refractive medium. As in the Letter, we use both numerical simulations and the analytical first Born approximation model to estimate excitation (and amplification) of positive and negative modes from the moving medium. We also consider a few simplified cases in which the Born approximation model can be solved analytically.\\

%\maketitle

\emph{\bf{Scattering from a superluminal medium:}}\\
A moving $\delta n$ with superluminal speeds may be obtained using Bessel pulses such as those used in Ref.~[1]. Here we show that our numerical and Born approximation models capture the essential features of the measured emission using Bessel pulse in fused silica glass [1,2]. \\
We first note that in diamond, if $v>c$ the dispersion curves both inside and outside the $\delta n$ resemble the curve for $\delta n_\textrm{max}$ in Fig.~1(f) of the Letter. so there is only one intersection point $\omega^\prime=\omega^\prime_\textrm{IN}$. Therefore only one mode (NRR) will be excited and there will be no amplification. We recall  that we are considering only forward propagating modes under the assumption that any interaction with backward modes, although in principle possible, will be extremely weak and inefficient if compared to any emission that will be observed when forward mode interaction is allowed. In the absence of coupling between forward modes, naturally coupling between forward and backward modes could be the only remaining interaction but will nevertheless be extremely weak and is therefore neglected in this work. \\
Conversely, in fused silica glass the dispersion relation exhibits a resonance in the infrared that allows coupling to two output RR and NRR modes or between two $\omega^\prime=0$ modes. The dispersion relations for fused silica relative to recent experiments using such Bessel-induced superluminal $\delta n$ are shown in Fig.~\ref{fig4}(a). If the probe input pulse starts from {\emph{inside}} the $\delta n$, then the pulse will start to lag behind, exit the $\delta n$ and, when traversing the $\delta n$ gradient it will excite RR and NRR output modes. Figure~\ref{fig4}(b) shows the evolution of the probe pulse as it interacts with the superluminal $\delta n$. Upon interacting with the gradient (grey shaded region) two modes are emitted. If we look at the spectrum evolution in Fig.~\ref{fig4}(c), surprisingly we see that there is complete conversion from the input mode to output RR and NRR modes even if there is no blocking horizon. The frequencies of these modes match perfectly with the predicted modes from the dispersion relation (dashed lines). 
\begin{figure}[t]
\centering
\includegraphics[width=8.5cm]{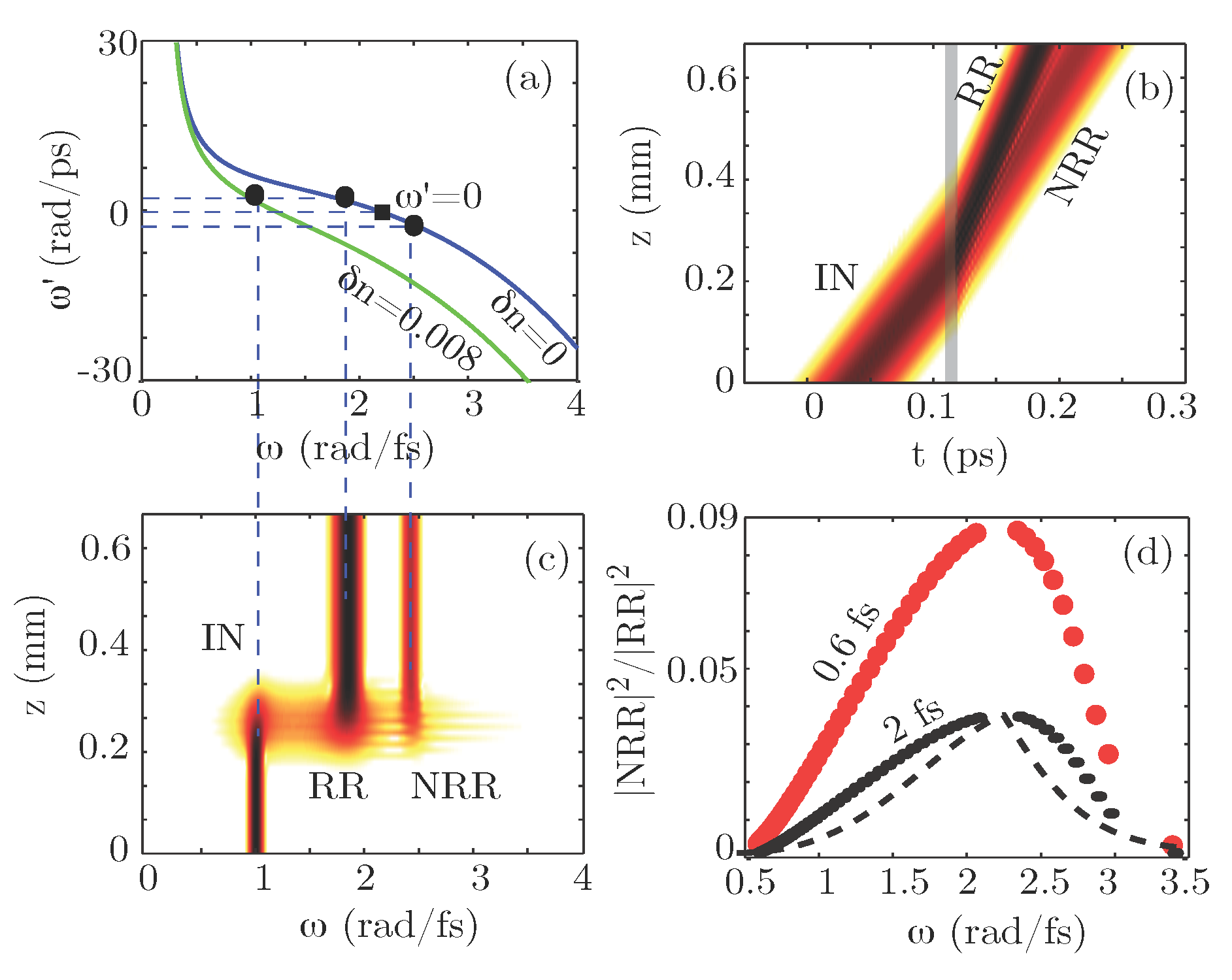}
\caption{Numerical simulation of a superluminal $\delta n$ in fused silica ($v=2.064\times10^8$ m/s). (a) dispersion relations outside the $\delta n$ and inside at $\delta n_\textrm{max}=0.008$.  (b) An example, of the temporal profile evolution for a 1.8 $\mu$m input wavelength pulse interacting with the superluminal $\delta n$. The input pulse is made to start from {\emph inside} the $\delta n$. (c) Same as in (b), showing the spectral evolution of the pulse. (d) $r=|NRR|^2/|RR|^2$ for two different $\delta n$ rise times as indicated in the figure. $r$ is peaked at the predicted comoving zero frequency, indicated with a fllled square in (a). Dashed curve - result from Born approximation calculation.}
\label{fig4}
\end{figure}
If we repeat the procedure outlined in the Letter, i.e. we repeat the simulations for varying input wavelength and monitor $r=|NRR|^2/|RR|^2$, we obtain the results in Fig.~\ref{fig4}(d) that shows curves for two different $\delta n$ rise times (0.6 fs and 2 fs). The dashed curve shows the result obtained for the 2 fs gradient pulse using the Born approximation and is in good agreement with the numerics.  In all cases, we see that $r$ is peaked at the $\omega^\prime=0$ frequency point on the $\delta n=0$ dispersion curve [indicated with a solid square in (a)]. This indicates that spontaneous emission will have maximum gain at this ``zero-frequency'' mode, in agreement with measurements of spontaneous emission  reported in Ref.~[1]. \\%Other attempts  to theoretically describe the measured emission resorted to simplified models that either neglected the $\delta n$ propagation or modelled dispersion through a time-varying mass. 
Although a quantitative comparison with analogue Hawking emission is somewhat thwarted  by the absence of a blocking horizon, which in turn poses a difficulty in making a direct comparison with a Hawking temperature [3], we see that the gain peak at $\omega^\prime\sim0$ in fused silica glass is essentially of the same nature of that observed in the blocking (and non-blocking) horizon case analysed in diamond. However, the latter material does seem more appropriate for future experiments aimed at observing and studying a faithful analogue of Hawking emission.\\

\emph{\bf{Photon number calculation:}}\\
The photon numbers, denoted e.g. with $|RR|^2$ and $|NRR|^2$ where calculated as 
\beq
\int_{\textrm{RR,NRR}} \cfrac{|E(\omega)|^2}{\omega}d\omega
\eeq
where $\int_{\textrm{RR,NRR}}$ indicates integration over the extent of the RR or NRR spectral peaks. These quantities are of course only proportional to the actual photon numbers as they neglect some constants that in any case cancel out when normalising to the input photon number and in any case do not change the functional dependence on $\omega$ that is the main focus of this work. The normalised photon number is thus taken as
\beq
|RR|^2,|NRR|^2=\cfrac{\int_{\textrm{RR,NRR}} \cfrac{|E(\omega)|^2}{\omega}d\omega}{\int_{\textrm{IN}} \cfrac{|E(\omega)|^2}{\omega}d\omega}.
\eeq
We note that in a non-dispersive medium the energy density is simply $u\propto|E|^2$ but in a dispersive medium it is corrected by a frequency dependent term, i.e. $u\propto|E|^2(\varepsilon+\omega d\varepsilon/d\omega)$, where $\varepsilon=\varepsilon(\omega)$ is the frequency dependent dielectric constant of the medium [4]. We have verified that in our simulations this corrective factor does actually modify in any way the results and can be safely neglected. \\

\emph{\bf{Analytical solutions to the Born approximation model:}}\\
%Notation: We use $\omega_{NRR},\omega_{RR},\omega_{RR}$ for frequencies  in the lab frame and $\omega,k_{IN},k_P,k_N$ for frequencies and momenta in  the comoving frame.\\
In the limit as $\omega' \to 0$ in the comoving frame, we are able to show 
analytically the thermal behaviour of the ratio $|N|^2/|P|^2$. In the 
lab frame we have to compute 

\beq
\frac{\omega_{NRR}^2}{k_{NRR}^2}\frac{k_{RR}^2}{\omega_{RR}^2} 
\left| \frac{R(\omega_{NRR}+\omega_{IN})}{R(\omega_{RR}-\omega_{IN})} 
\right|^2
\eeq
where $R$ is the Fourier transform of the perturbation $\delta n$. 

\subsection{Gaussian perturbation}

We choose
\beq
\delta n=\delta n_{max} e^{-u^2/\sigma^2}.
\eeq
Then 
\beq
R(s)=\sqrt{\frac{\sigma^2}{2}} e^{-1/4 s^2 \sigma^2},
\eeq
which implies 
\beq
\frac{R(\omega_{NRR}+\omega_{IN})}{R(\omega_{RR}-\omega_{IN})}
=e^{-1/4 (\omega_{NRR}+\omega_{RR})(\omega_{NRR}-\omega_{RR}+2 \omega_{IN})
\sigma^2}.
\eeq
In the comoving frame, the latter quantity becomes 
\beq
e^{-1/4 \gamma^2 v^2 (k'_N+k'_P)(k'_N-k'_P+2 k'_{IN})
\sigma^2}.
\eeq
In the limit as $\omega'\to 0$, both $k'_P$ and $k'_N$ tend to a same positive 
value $k'_0$ (as can be seen from the dispersion curve in the comoving frame), 
whereas $k'_{IN}$ tends to zero linearly in $\omega'$: 
$k'_{IN}\sim \zeta \omega'$, where $\zeta$ is a constant.\\ 
As to the factor 
\beq
\frac{\omega_{NRR}^2}{k_{NRR}^2}\frac{k_{RR}^2}{\omega_{RR}^2},  
\eeq
it is easy to show that it tends to 1 as $\omega'\to 0$.\\

As a consequence, we get 
\beq
|N|^2/|P|^2 \sim e^{-\alpha \omega'} \quad \quad \hbox{as}\ \omega'\to 0,
\eeq
with 
\beq
\alpha = 2 \gamma^2 v^2 k'_0 \sigma^2 \zeta.
\eeq

%\newpage
\subsection{Step-like perturbation}

We consider the following profile: 
\beq
\delta n\propto \tanh (l u),
\eeq
where $l$ is introduced as a steepness factor. In this case, we 
get 
\beq
R(s)=i \frac{\sqrt{2\pi}}{l} \frac{1}{e^{\frac{\pi}{2l} s} -
e^{-\frac{\pi}{2l} s}}.
\eeq
Then the ratio
\beq
\frac{R(\omega_{NRR}+\omega_{IN})}{R(\omega_{RR}-\omega_{IN})}
\eeq
in the comoving frame becomes
\beq
\frac{e^{\frac{\pi}{2l} \gamma v (k'_P-k'_{IN})} -
e^{-\frac{\pi}{2l} \gamma v (k'_P-k'_{IN})}}{e^{\frac{\pi}{2l}
\gamma v (k'_N+k'_{IN}) } -
e^{-\frac{\pi}{2l} \gamma v (k'_N+k'_{IN})}}.
\eeq
As far as $k'_P-k'_{IN}$ and $k'_N+k'_{IN}$ are positive and finite, we 
assume to be allowed to consider as negligible the second term (exponential with negative 
argument) both in the numerator 
and in the denominator of the previous equation. Then we get
\beq
|N|^2/|P|^2 \sim e^{\frac{\pi}{l} \gamma v (k'_P-k'_N+2 k'_{IN})} 
\sim e^{-2\frac{\pi}{l} \gamma v \zeta \omega'}
\eeq
as $\omega'\to 0$.

\subsection{Another step-like perturbation}

We consider the following profile: 
\beq
\delta n\propto \arctan (q u),
\eeq
where $q>0$ is introduced as a steepness factor.
In this case, we 
get 
\beq
R(s)=i \sqrt{\frac{\pi}{2}} \frac{e^{-\frac{|s|}{q}}}{s}.
\eeq
Then the ratio
\beq
\frac{R(\omega_{NRR}+\omega_{IN})}{R(\omega_{RR}-\omega_{IN})}
\eeq
in the comoving frame becomes
\beq
e^{\gamma v (k'_P-k'_N-2 k'_{IN})} \frac{k'_P- k'_{IN}}{k'_N+ k'_{IN}},
\eeq
which in the limit as $\omega' \to 0$ becomes 
\beq
e^{-2 \gamma v \frac{1}{q}\zeta \omega'},
\eeq
where $\zeta=$const as in the previous examples. Then we get 
\beq
|N|^2/|P|^2 \sim e^{-4\frac{1}{q} \gamma v \zeta \omega'}
\eeq
as $\omega'\to 0$.\\

Regarding the case of supergaussian profiles $\delta n=\delta n_{max} e^{-u^{2m}/\sigma^{2m}}$ with $m>1$, the analytical solutions become somewhat involved and it is more advantageous to resort to numerical solutions. However, we can 
point out that as $m$ increases the trailing profile of the 
pulse  looks approaches a step-like function of the kind we studied in the previous examples, 
so we can safely expect an analogous behaviour as $\omega^\prime\to 0$. This is indeed confirmed by our numerical results and the calculations show here underline the generality of the result that indeed, is not limited to just one specific functional shape for $\delta n$.

\subsection{Concluding remarks on the use of diamond for the measurement of analogue Hawking radiation}
In our setting, a moving $\delta n$ generated by an intense laser pulse may be tuned by controlling the pump laser pulse propagation velocity and intensity so as to form a blocking horizon for incoming light modes and ideal conditions for future measurements of analogue Hawking emission.
One of the main difficulties in past measurements of analogue Hawking radiation in optical systems was related to the shape of the dispersion relation in fused silica that exhibits a resonance at infrared wavelengths [1]. This resonance implies that the low-frequency behaviour of the artifical horizon is severely distorted from the linear, low frequency dispersion that is advocated as a fundamental requisite for observing a true analogue to Hawking radiation [5].
Diamond is ideal in this sense as it does indeed exhibit a practically linear dispersion relation down to extremely low frequencies whilst at the same time retaining a high optical nonlinearity and transparency across the whole electromagnetic spectrum, thus enabling measurements using nonlinear optics combined with a Hawking-like analysis of the emission.\\
%The emission studied here is stimulated by a second laser pulse but may be likewise stimulated by vacuum modes and would then bear all the attributes of spontaneous analogue Hawking radiation. 

%%%%%%%%%%%%%%%%%%%%%%%%%%%%%%%%%%
%\begin{thebibliography}{99}

[1] F. Belgiorno, S.L. Cacciatori, M. Clerici, V. Gorini, L. Rizzi, G. Ortenzi, E. Rubino, V.G. Sala, D. Faccio, Phys. Rev. Lett., {\bf 105}, 203901 (2010).\\

[2] E. Rubino, F. Belgiorno, S.L. Cacciatori, M. Clerici, V. Gorini, G. Ortenzi, L. Rizzi, V.G. Sala, M. Kolesik and D. Faccio, New J. Phys., {\bf 13}, 085005 (2011).\\

[3] R. Sch\"utzhold, W. G. Unruh, Phys. Rev. Lett., {\bf 107}, 149401 (2011).\\

[4]  F.D. Nunes et al., J. Opt. Soc. Am. B, {\bf 28},  1544-1552 (2011).\\

[5] C. Barcel\'o, S. Liberati, M. Visser, Living Rev. Relativity, {\bf 14}, 3  (2011). \\

%%%%%%%%%%%%%%%%%%%%%%%%%%

%\end{thebibliography}

\end{document}